\documentclass{svproc}
\usepackage{url}
\usepackage{graphicx} 

\begin{document}
\mainmatter              
\title{Radial flow induced by inhomogeneous magnetic field in heavy ion collisions}
\titlerunning{Hamiltonian Mechanics}  
%
\author{M. Haddadi Moghaddam\inst{1,2} \and B. Azadegan\inst{2}
\and A. F. Kord\inst{2} \and W. M. Alberico\inst{1}}
\authorrunning{M. Haddadi Moghaddam et al.} 
%
\tocauthor{}
\institute{Department of Physics, University of Turin and INFN, Turin, Via P. Giuria 1, I-10125 Turin, Italy
\and
Department of Physics, Hakim Sabzevari University (HSU), P.O.Box 397, Sabzevar, Iran}

\maketitle              

\begin{abstract}
In this paper, we study the effects of an in-homogeneous magnetic field on the quark gluon plasma dynamics,
within the Magneto-Hydrodynamic framework. We have investigated the effect of an inhomogeneous external
magnetic field on the transverse expansion of in-viscid fluid created in high energy nuclear collisions. Transverse velocity
and energy density are modified by the presence of the magnetic field.
This effects can also influence the transverse momentum spectrum for particles at the freeze-out surface. In this context
we obtained an interesting comparison with data extracted from heavy-ion collisions.
\keywords{Heavy ions collision, Magneto-hydrodynamic}
\end{abstract}
\section{Introduction}
Collisions of two heavy nuclei at high energy produce a hot and dense fireball.
Quarks and gluons could reach the deconfined state, called quark gluon plasma (QGP),
in a very short time ($\sim$ 1fm/c) after the initial hard parton collisions of nuclei. recently a wide range of
studies has shown that relativistic heavy-ion collisions create  huge
magnetic field due to the relativistic motion of the colliding heavy ions carrying large positive
electric charge  (For more references and details refer to it \cite{Tuchin:2013apa}-\cite{Gursoy:2014aka}. ).

Recently, some efforts in numerical and analytical works have been made, based on the relativistic
magneto-hydrodynamic  (RMHD) setup, to describe
high energy heavy ion collisions (See, for example, \cite{pu}-\cite{Das}).
The aim of our work is to generalize the Bjorken model by considering an inhomogeneous external magnetic
field acting on the medium. We show that the presence of the magnetic field leads to non-zero radial flow \cite{Haddadi et al2019}.

\section{Relativistic magneto-hydrodynamic}
We deal with the case of an ideal non-resistive plasma, with vanishing electric field in the local
rest-frame ($e^\mu=0$), which is embedded in an external magnetic field ($b_\mu$)
\cite{geod}-\cite{an89}.
The energy momentum conservation equations read:
\begin{eqnarray}\label{eq:conserve}
d_\mu(T_{pl}^{\mu\nu}+T_{em}^{\mu\nu})=0,
\end{eqnarray}
where
\begin{eqnarray}
T_{pl}^{\mu\nu}&=&(\epsilon+P)u^\mu u^\nu+Pg^{\mu\nu}\\
T_{em}^{\mu\nu}&=&b^2 u^\mu u^\nu+\frac{1}{2}b^2 g^{\mu\nu}-b^\mu b^\nu .
\end{eqnarray}
In the above $g_{\mu\nu}$ is the metric tensor, $\epsilon$ and $P$ are the energy density and
pressure, respectively. Moreover $d_\mu$ is the covariant derivative.
And the four velocity is defined as
$$u_\mu=\gamma(1, \vec v),\ \gamma=\frac{1}{\sqrt{1-v^2}}$$
satisfying the condition $u^\mu u_\mu=-1$.

Canonically one takes projections of the equation $d_\mu(T_{pl}^{\mu\nu}+T_{em}^{\mu\nu})=0$
along the parallel and perpendicular directions to $u_\nu$,
 which gives:
\begin{eqnarray}
D(\epsilon+b^2/2)+(\epsilon+P+b^2)\Theta+u_\nu b^\mu(d_\mu b^\nu)&=&0,\label{eq:energycons}\\
(\epsilon+P+b^2)Du^\alpha+\nabla^\alpha(P+\frac{1}{2}b^2)-d_\mu (b^\mu b^\alpha)-u^\alpha
u_\nu d_\mu( b^\mu b^\nu)&=&0.\label{eq:momentumcons}
\end{eqnarray}
Notice that $\alpha$ should be a spacelike index. Moreover
\begin{eqnarray}
D=u^\mu d_\mu,\ \ \ , \Theta=d_\mu u^\mu,\ \ \ , \nabla^\alpha=\Delta^\alpha_{\ \nu}d^\nu.
\end{eqnarray}
\subsection{Induced radial flow in B-field}
We consider the medium  expands both
radially and along the beam axis, the only nonzero components of $u_\mu=(u_\tau, u_\perp, 0, 0)$ are $u_\tau$,
which describes the boost-invariant longitudinal expansion \cite{bj83},
and $u_\perp$, which describes the transverse expansion.
And we suppose the external magnetic field to be located in transverse plane as $b_\mu=(0, 0, b_\phi, 0)$.

We now seek the perturbation solution in the presence
of a weak external magnetic field  pointing along the $\phi$
direction in an inviscid fluid with infinite electrical conductivity:
\begin{eqnarray}
&&u_\mu=(1, \lambda^2u_\perp, 0, 0),
 b_\mu=(0, 0, \lambda b_\phi, 0),\ \ b^2\equiv b^\mu b_\mu\nonumber\\&&
\epsilon=\epsilon_0(\tau)+\lambda^2\epsilon_1(\tau, x_\perp),\ \epsilon_0(\tau)=\frac{\epsilon_c}{\tau^{4/3}}
\end{eqnarray}

In such setup, the conservation equations (\ref{eq:energycons}-\ref{eq:momentumcons})
 reduce to the following partial differential equations
\begin{eqnarray}
&&u_\perp-\tau^2\partial_\perp(\frac{u_\perp}{x_\perp})-\tau^2\partial_\perp^2u_\perp-\tau\partial_\tau u_\perp+3\tau^2\partial_\tau^2u_\perp\nonumber\\&&
-\frac{3\tau^{7/3}}{x_\perp\epsilon_c}b_\phi^2-\frac{3\tau^{7/3}}{4\epsilon_c}\partial_\perp
b_\phi^2-\frac{9\tau^{10/3}}{4x_\perp\epsilon_c}\partial_\tau b_\phi^2-\frac{3\tau^{10/3}}{4\epsilon_c}\partial_\perp\partial_\tau b_\phi^2=0.\label{inh}
\end{eqnarray}

For non-vanishing $b_\phi$ we assume a space-time profile of the magnetic field in central collisions in the form:
\begin{eqnarray}
b_\phi^2(\tau, x_\perp)=B_c^2 \tau^{-1}\sqrt{\alpha} x_\perp e^{-\alpha x_\perp^2}.
\label{magnetic}
\end{eqnarray}
We see that the magnitude of $b_\phi$ is zero at $x_\perp=0$. Finally, one can find the solutions for transverse velocity $u_\perp$ and correspondingly modified energy density $\epsilon_1$; for more details refer to \cite{Haddadi et al2019}.

\section{Particle transverse momentum spectrum}
From the local equilibrium hadron distribution the transverse
spectrum is calculated via the Cooper-Frye  formula in the freeze out surface
\begin{eqnarray}
S=E\frac{d^3N}{dp^3}&=&\frac{g_i}{2\pi^2}\int_0^{x_f}\  x_\perp\ \tau_f(x_\perp)\ dx_\perp\
\Big[m_T K_1(\frac{m_Tu_\tau}{T_f})I_0(\frac{m_Tu_\perp}{T_f})\nonumber\\&&
+p_T R_f K_0(\frac{m_Tu_\tau}{T_f})I_1(\frac{m_Tu_\perp}{T_f})\Big]
\label{spectrum}
\end{eqnarray}
Where $\tau_f(x_\perp)$ is the solution of the $T(\tau_f, x_\perp)=T_f$ and the degeneracy is $g_i=2$
for both the pions and the protons. The above integral over $x_\perp$ on the freeze-out surface is evaluated numerically.

The spectrum Eq.~(\ref{spectrum}) is illustrated in the following figures for three different
values of the freeze out temperature (140, 150 and 160 MeV) and compared with experimental results obtained at PHENIX \cite{phenix}.
in central collisions.
Our proton spectrum appear to underestimate the experimental data, except at low $p_T$, but their behavior with $p_T$has the correct trend of
a monotonically decrease. The pion spectrum, instead, appears in fair agreement with the experimental results, which are very close to the
theoretical curves. This is an indication that hadrons with different masses have different sensitivities to the underlying hydrodynamic flow
and to the electromagnetic fields. Indeed, the difference between the charge-dependent flow of light pions and heavy protons might arise
because the former are more affected by the weak magnetic field than the heavy protons \cite{gursoy2018}.

For comparison, we also show the results obtained by Gubser \cite{gubser}, which appear to be more flat and typically overestimate the experiment.
We also notice that, for the proton case, the highest value of the freeze out temperature we employed (as suggested, e.g. in Ref.~\cite{Ratti})
slightly brings (for protons) the calculation closer to the experimental data; however it also shows a kind of saturation phenomenon
and points to the need of including other effects not considered in the present work.

\begin{figure}[ht]
\begin{minipage}[b]{0.45\linewidth}
\centering
\includegraphics[width=1\textwidth]{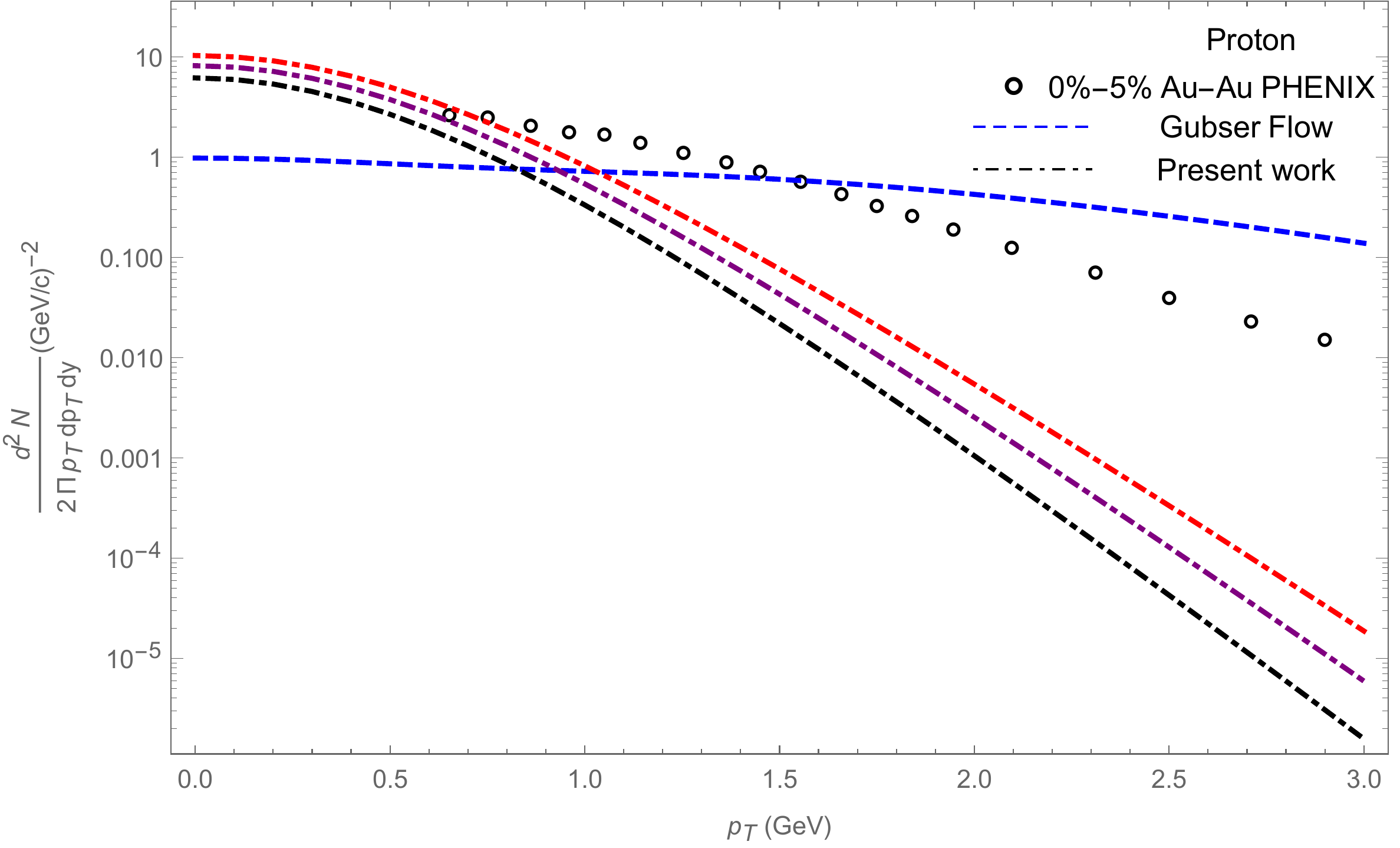}
\end{minipage}
\hspace{0.5cm}
\begin{minipage}[b]{0.45\linewidth}
\centering
\includegraphics[width=1\textwidth]{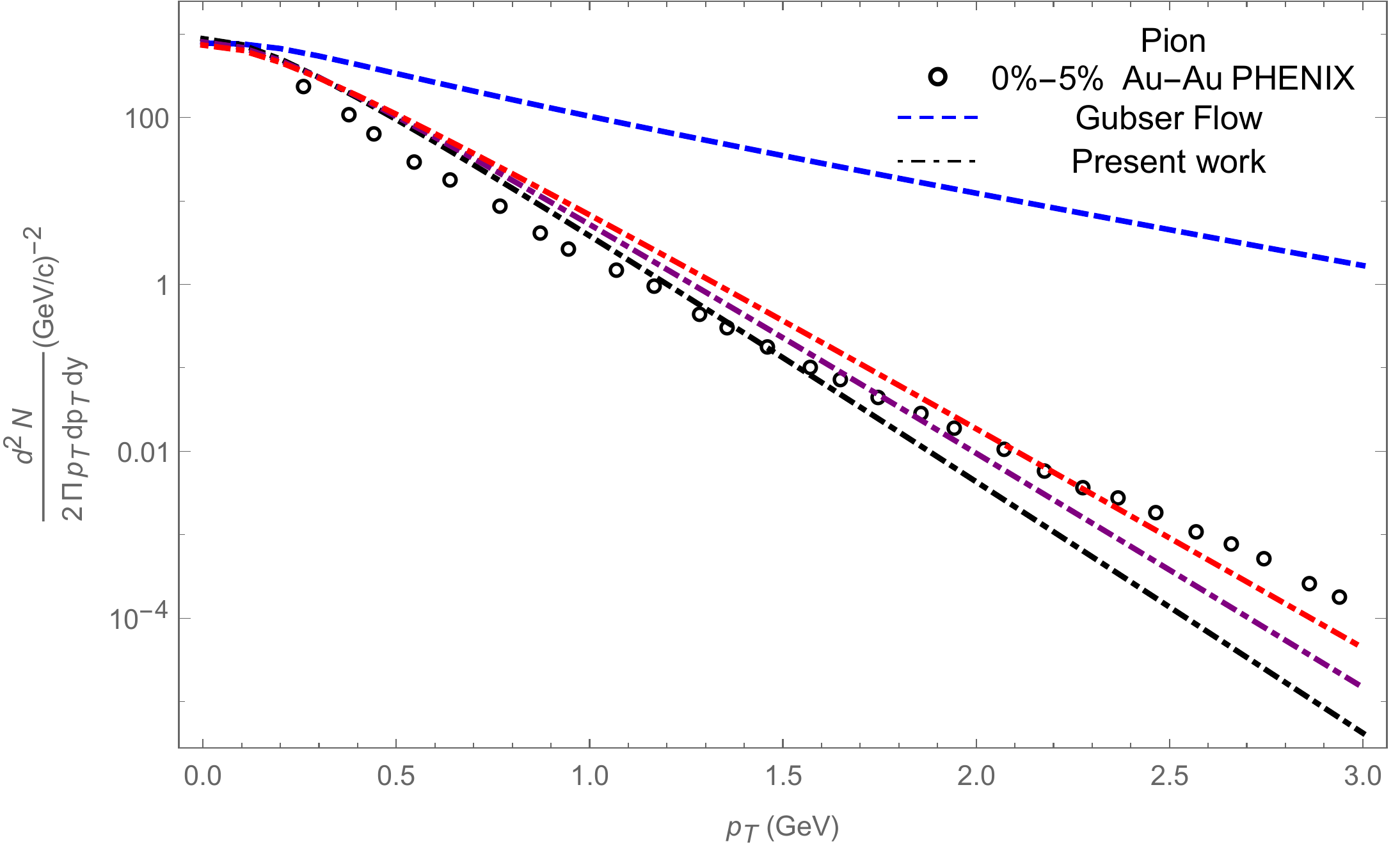}
\end{minipage}
\caption{ Particles transverse momentum spectrum from central Au-Au collisions: black, purple and red lines correspond to a
freeze out temperature of 140, 150 and 160 MeV, respectively. Circles: PHENIX data~\cite{phenix}. }
\end{figure}

%

\end{document}